\documentclass[]{aa}
\usepackage{txfonts}
\usepackage{graphicx}

\usepackage{natbib}
\bibpunct{(}{)}{;}{a}{}{,}

\usepackage{floatflt}
\begin{document}
\title{Self--consistent treatment of dynamics and chemistry in the winds from carbon-rich 
AGB stars. I. Tests of the equilibrium and kinetic chemical codes.}
\author{M. Pu{\l}ecka\inst{1}
   \and M.R. Schmidt\inst{1}
   \and V.I. Shematovich\inst{2}
   \and R. Szczerba\inst{1}
         }
\institute{N. Copernicus Astronomical Center, Rabia\'{n}ska 8,
               87--100 Toru\'n, Poland
        \and
           Institute of Astronomy, Russian Academy of Sciences, Pyatnitskaya 48,
           Moscow, Russian Federation
                    }
\date{Received; Accepted }
\abstract
   {}
   {The main aim of the paper was performing test of our (chemical and kinetic) codes, 
which will be used during self-consistent modelling of dynamics and chemistry in the 
winds from C-rich AGB stars.}
   {We use the thermodynamical equilibrium code to test the different databases 
of dissociation constants.
We also calculate the equilibrium content of the gas
using the kinetic code, which includes the chemical network of 
neutral--neutral reactions studied by Willacy \& Cherchneff\,(1998). The
influence of reaction rates updated using the  UMIST database 
for Astrochemistry 2005 (UDFA05), was tested.
}
   { The local thermodynamical equilibrium calculations have shown that 
the NIST database reproduces fairly well equilibrium concentrations of 
Willacy \& Cherchneff\,(1998), while agreement in case of Tsuji\,(1973) dissociation
constants is much worse.

 The most important finding is that the steady state solution obtained with the kinetic 
code for reaction network of Willacy \& Cherchneff\,(1998) is different from the 
thermodynamical equilibrium solution. In particular, CN and C$_2$, which are important 
opacity sources are underabundant relative to thermodynamical equilibrium,  while O-bearing 
molecules (like SiO, H$_{2}$O, and OH) are overabundant. After updating the reaction rates 
by data from the UDFA05 database consistency in O-bearing species becomes much better, 
however the disagreement in C-bearing species is still present.
}
 {}

\titlerunning{Test of the chemical codes.}
\authorrunning{Pulecka et al.}

\maketitle

\section{Introduction}

Asymptotic Giant Branch (AGB) stars develop a molecular dusty wind, driven by
radiation pressure on dust grains condensating in the warm molecular layers.
The new formed molecules and dust are deposited then into interstellar medium (ISM)
and contribute to the global matter budget of galaxies. Therefore, the
determination of molecular and dust contents in the AGB winds is important for
understanding of their further re-processing in the ISM before, eventually, matter
ends up in a new star forming regions. 

Chemical evolution of gas around AGB stars takes place at such different
conditions like thermodynamical equilibrium in stellar photospheres, non--equilibrium
(including possible shock--induced) chemistry, dust formation and photon dominated 
chemistry in circumstellar envelopes. 

Observations of circumstellar envelopes at infrared and millimetre range
demonstrate that they are very rich in molecules. In the case of the well--studied
C--rich AGB star IRC+10\,216 the number of identified molecules exceeded 60
\citep{O06}. The majority of the observed molecules are carbon--chain 
molecules such as the cyanopolyynes, the hydrocarbon radicals and carbenes, as well
as organo--sulphur and organo--silicon molecules. Interferometric and single dish
observations show that carbon--chain molecules are distributed in the hollow shells
around central star. Such distribution has been interpreted as arising
from photochemistry of parent molecules (e.g. C$_2$H$_2$ and HCN) flowing
away from the star \citep[see e.g. reviews by][]{G96,M03}.
The new space and ground--based projects (e.g. Stratospheric 
Observatory for Infrared Astronomy -- SOFIA, The Atacama Large Milimeter Array -- ALMA 
or Herschel Space Observatory) will allow us to investigate the circumstellar
molecular inventory in much more details by using, not yet fully exploited, 
sub--millimetre range. 

With the aim to prepare tools for analysing the new sub--millimetre observations
of AGB stars we are in a process of developing a self--consistent treatment of
dynamics and chemistry in the circumstellar envelopes. The first effort to build 
the chemical network of reactions for conditions close to thermodynamical equilibrium 
(high density and temperature) in C--rich environment was done by 
\citet{W98} (hereafter WC98). Since the considered densities are high and 
effective temperature of the star is low, only neutral--neutral bimolecular and termolecular 
reactions were considered. WC98 investigated chemistry of sulphur 
and silicon in the inner wind of IRC+10216. Their non--equilibrium calculations started from 
chemical composition derived at the local thermodynamical equilibrium (LTE) conditions.
It was found that the parent molecules C$_2$H$_2$ 
and CO are unaffected by the shocks, but the abundances of daughter molecules can be 
significantly altered by the shock--driven chemistry. Their model indicated that molecule 
production by the shock--driven chemistry close to the photosphere can be significantly 
more important than the LTE chemistry. Recently, Cherchneff\,(2006) investigated the chemistry 
and composition of the quasi-static molecular layers of AGB stars using the same 
semi-analytical model of shock dynamics as in WC98. 

In modelling of stellar atmospheres it is commonly assumed that
conditions of local thermodynamical equilibrium prevails. Molecular content is then
defined by dissociation constants. In the outflow, in the presence of shocks,
kinetic equations must be solved for some chemical network. When local thermodynamical 
equilibrium is valid, at some position in the stellar atmosphere, both approaches 
should produce the same molecular concentrations. If not, then an
inconsistency will appear, darkening the interpretation of obtained results.
The aim of our paper is to clarify this point for future application of available
chemical networks in the non-equilibrium computations of chemistry in the inner layers of 
carbon-rich outflows.

In this paper we present our steady--state chemical code and test (some of) the available 
chemical databases using for comparison work of WC98. In the next paper (Schmidt et al. 
in preparation) we will discuss our approach to self--consistent treatment of 
dynamics and chemistry. The first preliminary results of such computations were 
presented by \citet{P05} (using the TITAN hydrodynamical code) and 
\citet{P205} (using the FLASH hydrodynamical code). 

\section{Test of chemical codes: The thermodynamical equilibrium}

In the stellar photospheres the dynamical timescales are much longer than
the timescales of molecular formation \citep{G96}. Therefore, to
determine local molecular contents it is sufficient to apply the thermodynamical equilibrium (TE)
conditions.
\begin{figure*}[!ht]
\centering
\resizebox{\hsize}{!}{\includegraphics[angle=90]{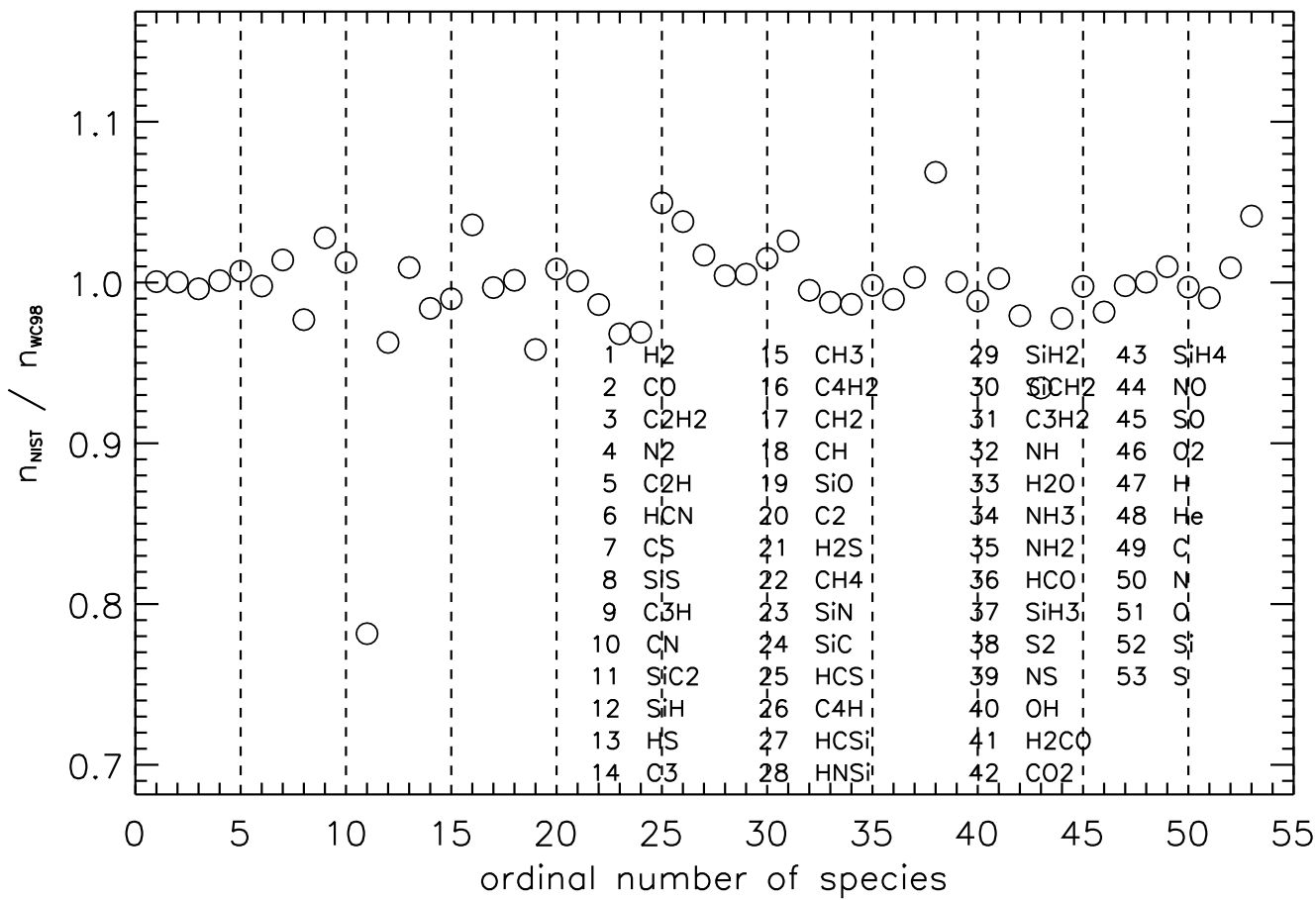}}
\caption{The ratio of NIST and WC98 equilibrium concentrations. The order of species
(except of elements, which are placed at the right side of the figure) follows decreasing
concentrations of WC98.}
\label{err_nist}
\end{figure*}
To compute molecular composition in the local TE conditions one needs to know the
temperature ($T$), the total gas number density ($n$), or equivalently the total
gas pressure ($p$), and the initial elemental abundances. We have constructed such
TE code and present results of the performed tests.

\subsection{The thermodynamical equilibrium code}
\label{TEcode}

The equilibrium chemistry code is based on the Russell's \citep{R34}
approach. This method uses the fictitious pressure $p_{\rm f}({\rm X})$, which is the
pressure exerted by element X if all the gas were in the neutral form of the atomic species X.
Hence the fictitious pressure of element X is given by:
\begin{equation}
p_{\rm f}{\rm (X)}= p_{\rm X} + \sum_{\rm Y} \alpha_{\rm X,Y} p_{\rm Y},
\end{equation}
where: $p_{\rm Y}$ -- partial pressure\footnote{The partial pressure of species Y is
the pressure of the gas if no other species were presented in the medium.}
of molecular species Y, which contains element X,
$\alpha_{\rm X,Y}$ -- the stoichiometric coefficient indicating the number of element X
involved in molecule Y. For example, in the case of hydrogen we have:
\begin{equation}
p_{\rm f}({\rm H}) =p_{\rm H}+2 p_{\rm H_2} + p_{\rm CH} + 2 p_{\rm CH_2} + ...
\end{equation}
(plus all examined hydrogen--bearing molecules).

Partial pressure of compound X
is related to its concentration $n_{\rm X}$ [cm$^{-3}$] by the ideal gas law
$p_{\rm X}\,=\,n_{\rm X}\,k\,T$ [dyn cm$^{-2}$], where $k$ -- Boltzmann constant
[erg/K], and $T$ -- temperature [K]. Using the partial pressures, the dissociation
constant (K$_{\rm p}$) of molecule AB (AB~$\rightarrow$~A~+~B) can be expressed as:
\begin{equation}
K_{\rm p}({\rm AB})
=\frac{p_{\rm A} p_{\rm B}}{p_{\rm AB}}
= \frac{n_{\rm A}\, kT\, n_{\rm B}\,kT}{n_{\rm AB}\,kT}
= \frac{n_{\rm A}\,n_{\rm B}}{{n_{\rm AB}}}\,kT,
\label{kp}
\end{equation}
where: $p_{\rm A}$, $p_{\rm B}$, $p_{\rm AB}$ -- partial pressures
of atoms A, B, and molecule AB, $n_{\rm A}$, $n_{\rm B}$, $n_{\rm AB}$ -- their
concentrations. 
Therefore by employing the Eq.~\ref{kp}, the fictitious pressure of hydrogen can be 
rewritten to:
\begin{equation}
p_{\rm f}{(\rm H)} = p_{\rm H}+2 \frac{p^2_{\rm H}}{K_{\rm p}{(\rm H_2)}} +
\frac{p_{\rm H} p_{\rm C}}{K_{\rm p}{\rm(CH)}} + 2 \frac{p^2_{\rm H} p_{\rm C}}{
K_{\rm p}{\rm (CH_2)}} +...
\label{pf}
\end{equation}
 Note, that dissociation constants can be determined from the differences of the corresponding Gibbs free 
energies\footnote{Here the sign $^\circ$ marks conditions when temperature is 273.15 K and
the absolute pressure is 1 atm (i.e. 1.01325 bar, 1.01325$\times$ 10$^6$ [dyn cm$^{-2}$],
101.325 [kPa]). 
($\Delta G^{\circ}$) of products (A, B) and reactant (AB) according to:
\begin{equation}
K_{\rm p} (AB)= C^{\Delta \nu}\,\exp{\left(-\frac{\Delta G^{\circ}}{R\,T}\right)},
\label{deltaG}
\end{equation}
where: $R=8.32441\times10^7$ [erg K$^{-1}$ mol$^{-1}$]
is the ideal gas constant, C = 1.01325$\times$ 10$^6$ [dyn cm$^{-2}$] and
$\Delta \nu$ is the stoichiometric factor of a given dissociation reaction.}
Since, the fictitious pressure reflects the conservation of mass for element X, then
$p_{\rm f}$(X) for any other atomic species X can be expressed in terms of the
fictitious pressure of hydrogen $p_{\rm f}$(H):
\begin{equation}
p_{\rm f}({\rm X}) = A({\rm X}) p_{\rm f}({\rm H}),
\label{Apf}
\end{equation}
where: $A$(X) = $\frac{n_{\rm X}}{n_{\rm H}}$ is abundance of element X relative to
hydrogen.
\begin{figure*}[!ht]
\centering
\resizebox{\hsize}{!}{\includegraphics[angle=90]{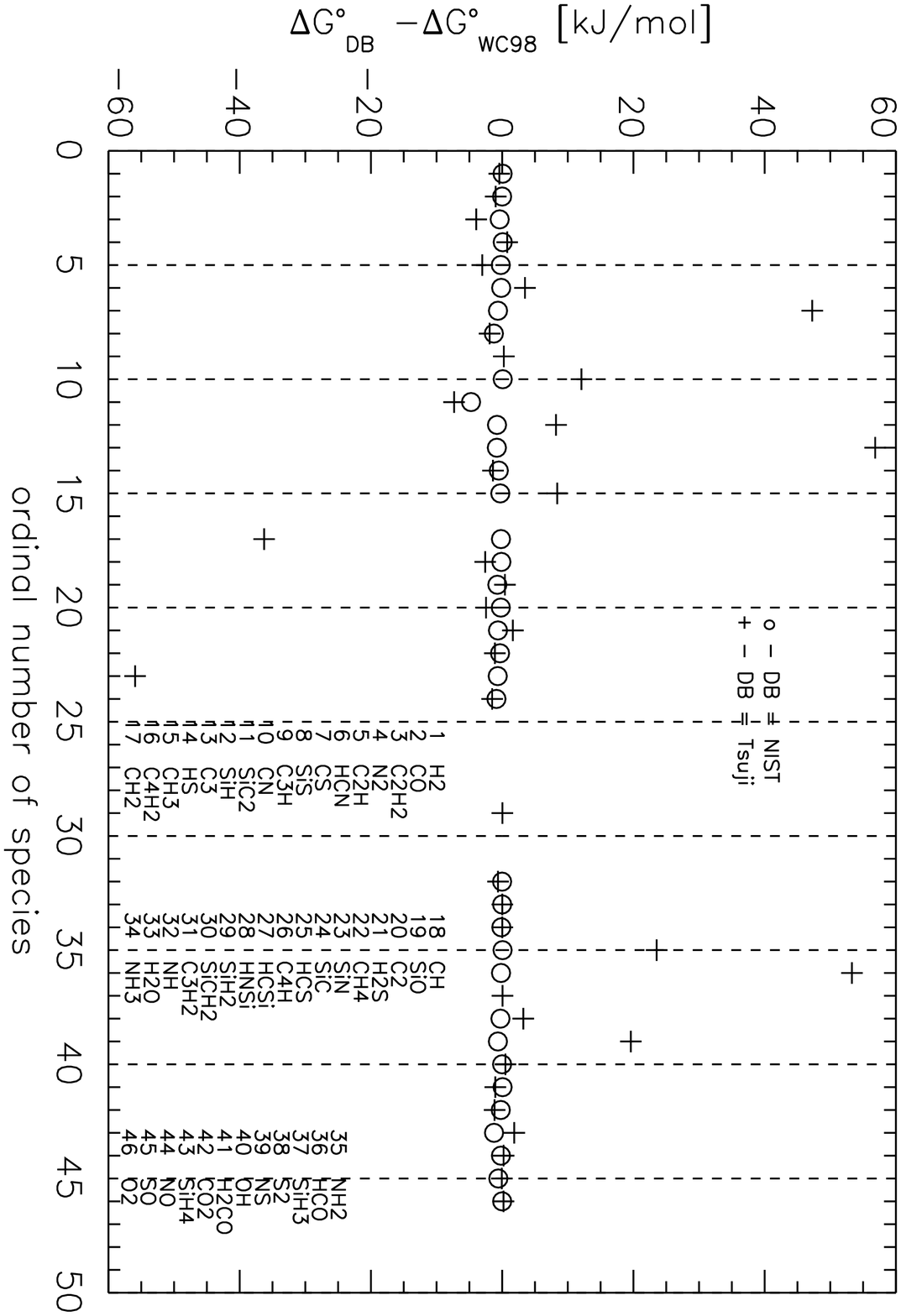}}
\caption{The differences of $\Delta G^{\circ}$ between database (DB: DB = NIST
 or Tsuji) and WC98. Results are marked by circles for the NIST database 
and by pluses for the Tsuji database. The order of species is same as in Fig.\,\ref{err_nist}}
\label{err_gibbs}
\end{figure*}

To close this set of nonlinear equations for fictitious pressures
we need only the equation for the total gas pressure, which is obtained by
summing the partial pressures of all examined species.
This final set of equations is solved using the Netwon-Raphson`s method.
The solution gives the partial pressure of atoms,
which are used then to calculate molecular partial pressures (i.e. in fact
their concentrations) by employing Eq.~\ref{kp}.

\subsection{The equilibrium case of WC98}
\label{TEWC98}
In a first test of our thermodynamical equilibrium code we aimed to reproduce the thermodynamical
equilibrium concentrations of species given in Table\,3 of WC98. This table contains equilibrium
abundances determined at distance of 1.2\,R$_{*}$
and physical conditions specified in the 1$^{st}$ entry of their Table\,2
($T$\,=\,2062 [K] and $n$\,=\,3.68$\times$10$^{14}$ [cm$^{-3}$] what corresponds to
$p$\,=\,1.05$\times$10$^2$ [dyn~cm$^{-2}$]). To reproduce the equilibrium molecular
content at these conditions we need to know the initial elemental abundances and
the dissociation constants.

The initial elemental abundances were determined by summing abundances of all species
(see Table\,3 of WC98) containing a given element. The obtained values are given in 
Col. 2 of Table~\ref{elemAbun}, where abundance of element X is given by
$\epsilon({\rm X}) = log \frac{n_{\rm X}}{n_{\rm H}} + 12$, and $n_{\rm X}$
is concentration of element X. For comparison, the solar abundances
of H, He, C, O, N, Si, and S from \citet{G98} (hereafter GS98)
and \citet{A05} (hereafter A05) are shown in Cols. 3 and 4,
respectively. The initial element abundances assumed by WC98 are essentially
solar ones except of C/O ratio, which was assumed to be about 1.5.

\begin{table} [!ht]
\centering
\caption[short]{\normalsize{The initial abundances of elements:
$\epsilon({\rm X}) = log \frac{n_{\rm X}}{n_{\rm H}} + 12$.}}
\vspace{0.01cm}
\begin{tabular}{*{4}{c}}
\hline
\hline
element & WC98 & GS98 & A05 \\
 (1) & (2) & (3) & (4) \\ \hline
 H & 12.00 & 12.00 & 12.00 \\
 He & 10.99 & 10.93 & 10.93 \\
 C & 8.98 & 8.56 & 8.39 \\
 N & 8.06 & 8.05 & 7.78 \\
 O & 8.79 & 8.93 & 8.66 \\
 Si & 7.49 & 7.55 & 7.51 \\
 S & 7.19 & 7.33 & 7.14 \\ \hline
\end{tabular}
\label{elemAbun}
\end{table}

The dissociation constants were obtained using species
concentrations given in Table 3 of WC98 by employing Eq.~\ref{kp}.
Note that the dissociation constants derived
in such a way are valid {\it only} for the conditions (temperature)
specified above. As expected,  with such $K_{\rm p}$'s and initial 
elemental abundances given in column 1 of Table\,\ref{elemAbun}, 
the obtained equilibrium concentrations are in a very good agreement  
with values given in Table 3 of WC98. However, after discussion
with K. Willacy, we realized that there are some misprints for the concentrations of
C$_2$ and SiH$_4$ in Table 3 of WC98.
Therefore, during further computation we have implemented the original $\Delta G^{\circ}$'s
(kindly provided by K. Willacy) from work of WC98. 
These $\Delta G^{\circ}$'s are collected now in Table~\ref{stale} (Cols. 3 and 8), 
and were used to determine equilibrium species concentrations which
will be referred hereafter as WC98 concentrations. The molecules listed in 
this table (with ordinal numbers in Cols. 1 and 6) are organised in order of decreasing 
WC98 concentrations.

\subsection{Comparison of TE concentrations based on WC98, NIST and Tsuji databases}
As~was~described~in~Sect.~\ref{TEWC98}~determination~of TE concentrations at specified 
conditions requires knowledge of $K_{\rm p}$'s  (or equivalently $\Delta G^{\circ}$'s). 
The most comprehensive
database, which is publicly available and allows to compute dissociation constants is that
collected by the National Institute of Standards and Technology (NIST).
The NIST chemistry WebBook\footnote{http://webbook.nist.gov/chemistry/}
provides thermochemical data, like enthalpy of formation ($H^{\circ}$) and molar
entropy ($S^{\circ}$) for different temperatures at standard
conditions. These quantities allow determination of the
Gibbs free energy ($G^{\circ}$) for all examined species according to:
\begin{equation}
G^{\circ} = H^{\circ} - T\ S^{\circ}.
\label{gibbs}
\end{equation}
Thus, we can determine the difference of Gibbs free energies ($\Delta G^{\circ}$) for 
products and reactants, of the direct dissociation of molecule into single atoms, and 
use them to find the dissociation constants from Eq.~\ref{deltaG}.

 We computed concentrations of species according to the method described in
Sect.\,\ref{TEcode}, using the dissociation constants obtained from the NIST 
database for $T$ = 2062 [K].  The NIST database does not contain thermochemical data for some 
molecules: C$_3$H -- \#9 (number appearing after symbol \# is the ordinal number of a given 
molecule used in Fig.\,\ref{err_nist}), 
C$_4$H$_2$ -- \#16, HCS -- \#25, C$_4$H -- \#26,  HCSi -- \#27, HNSi -- \#28, SiH$_2$ -- \#29, 
SiCH$_2$ -- \#30,  C$_3$H$_2$ -- \#31, SiH$_3$ -- \#37, so we have used $\Delta G^{\circ}$'s 
from WC98, instead. The comparison between NIST and WC98 concentrations is shown in 
Fig.\,\ref{err_nist}. 
The order of species (except of elements, which are placed at the right side of the figure) 
follows decreasing concentrations of WC98. The consistency is fairly good (almost all results 
agree within about 7 \%, with the exception of SiC$_2$ -- \#11 concentration, 
for which difference is about 22\,\%.

\begin{table*} [!ht]
\centering
\caption[short]{\normalsize{$\Delta G^{\circ}$ in [kJ/mol] for $T$\,=\,2062\,K
and $p$ = 1 atm in case of WC98, NIST, and \citet{T73}.}}
\vspace{0.5cm}
{\scriptsize
\begin{tabular}{*{10}{c}}
\hline
\hline
\# & species & WC98 & NIST & TSUJI & \# & species & WC98 & NIST & TSUJI \\
(1)& (2)                   & (3) & (4) & (5) & (6)& (7) & (8) & (9) & (10) \\ \hline
 1 & H$_2$            &     207.6 & 206.1 &   205.6& 24 & SiC              &  215.4 & 214.5 &   213.9 \\
 2 & CO               &     801.6 & 801.6 &   800.6& 25 & HCS              &  463.0 &       &         \\
 3 & C$_2$H$_2$       &     878.4 & 878.0 &   874.4& 26 & C$_4$H           & 1279.8 &       &         \\
 4 & N$_2$            &     684.3 & 684.4 &   685.1& 27 & HCSi             &  386.0 &       &         \\
 5 & C$_2$H           &     676.0 & 675.8 &   673.0& 28 & HNSi             &  476.7 &       &         \\
 6 & HCN              &     770.0 & 769.8 &   773.5& 29 & SiH$_2$          &  150.4 &       &   150.4 \\
 7 & CS               &     453.8 & 453.2 &   501.1& 30 & SiCH$_2$         &  555.8 &       &         \\
 8 & SiS              &     372.7 & 371.4 &   370.8& 31 & C$_3$H$_2$       & 1038.2 &       &         \\
 9 & C$_3$H           &     1049.7&       &  1050.0& 32 & NH               &  109.9 & 109.9 &   109.3 \\
10 & CN               &     507.0 & 507.1 &   519.1& 33 & H$_2$O           &  455.4 & 455.4 &   455.4 \\
11 & SiC$_2$          &     740.9 & 736.2 &   733.6& 34 & NH$_3$           &  464.7 & 464.6 &   464.7 \\
12 & SiH              &     94.8  & 94.0  &   103.0& 35 & NH$_2$           &  277.1 & 277.2 &   300.6 \\
13 & C$_3$            &     794.3 & 793.5 &   851.1& 36 & HCO              &  662.7 & 662.5 &   716.0 \\
14 & HS               &     152.5 & 152.0 &   151.1& 37 & SiH$_3$          &  280.9 &       &   281.0 \\
15 & CH$_3$           &     516.5 & 516.2 &   524.9& 38 & S$_2$            &  183.7 & 183.4 &   186.9 \\
16 & C$_4$H$_2$       &     1550.3 & &             & 39 & NS               &  262.4 & 261.7 &   282.0 \\
17 & CH$_2$           &     318.8 & 318.6 &   282.5& 40 & OH               &  212.1 & 212.1 &   212.6 \\
18 & CH               &     131.7 & 131.6 &   129.1& 41 & H$_2$CO          &  763.3 & 763.4 &   762.3 \\
19 & SiO              &     537.6 & 536.9 &   538.0& 42 & CO$_2$           &  1024.3& 1024.1&  1023.1  \\
20 & C$_2$            &     346.4 & 346.2 &   343.9& 43 & SiH$_4$          &  335.1 & 333.9 &   336.9 \\
21 & H$_2$S           &     286.8 & 286.2 &   288.4& 44 & NO               &  395.5 & 395.3 &   395.7 \\
22 & CH$_4$           &     667.7 & 667.4 &   666.6& 45 & SO               &  279.3 & 278.7 &   279.2 \\
23 & SiN              &     317.3 & 316.6 &   261.4& 46 & O$_2$            &  234.8 & 234.8 &   235.0 \\
\hline
\end{tabular}
}
\label{stale}
\end{table*}
From chemical point of view a better illustration of discrepancies between  different
databases of thermochemical data is comparison between $\Delta G^{\circ}$'s. Therefore,
in Cols. 4 and 9 of Table\,\ref{stale} we provide the $\Delta G^{\circ}$'s derived
from the NIST database. We left empty spaces for species with missing data in that
database. In Fig.\,\ref{err_gibbs} 
we plotted the differences between $\Delta G^{\circ}$' from NIST and WC98 (circles) for each 
species. 
Again, the order of species (except of elements, which are placed at the right side of the figure) 
follows decreasing concentrations of WC98. For molecules without data in the NIST database 
there are no corresponding symbols in Fig.\,\ref{err_gibbs}. As one can see energies agrees 
within about 2 [kJ mol$^{-1}$] with exception of SiC$_2$ for which discrepancy 
is about 5 [kJ mol$^{-1}$].
\begin{figure}[!bt]
\centering
\resizebox{\hsize}{!}{\includegraphics[angle=90]{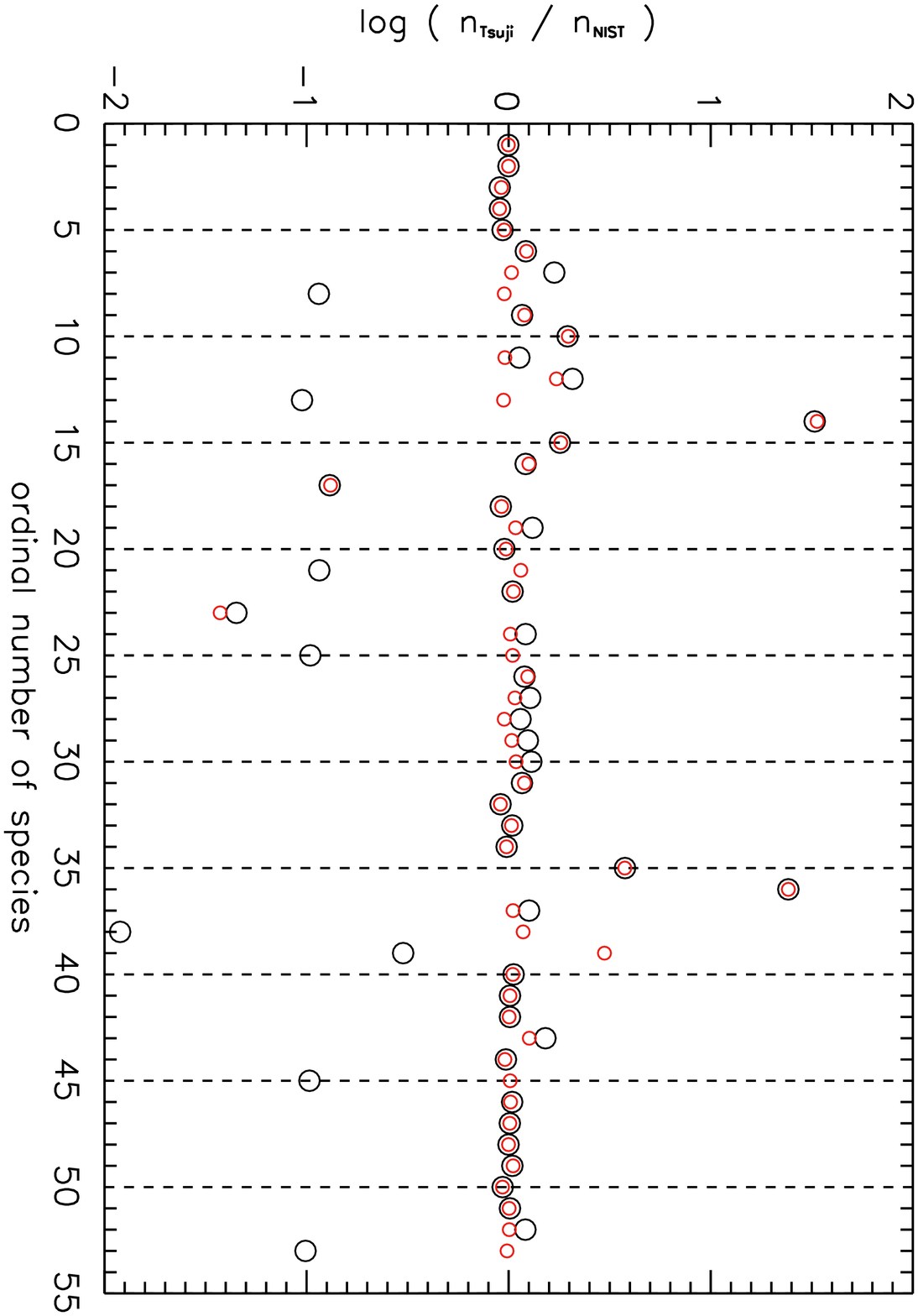}}
\caption{The ratio between Tsuji and NIST equilibrium concentrations at conditions 
specified in
Sect.\,\ref{TEWC98} -- black circles. Red circles mark the ratio, when
$\Delta G^{\circ}$ for CS in the Tsuji database was replaced by the value from WC98.
The order of species is same as in Fig.\,\ref{err_nist}}
\label{err_tsCS}
\end{figure}
\begin{figure}[!bt]
\centering
\resizebox{\hsize}{!}{\includegraphics[angle=90]{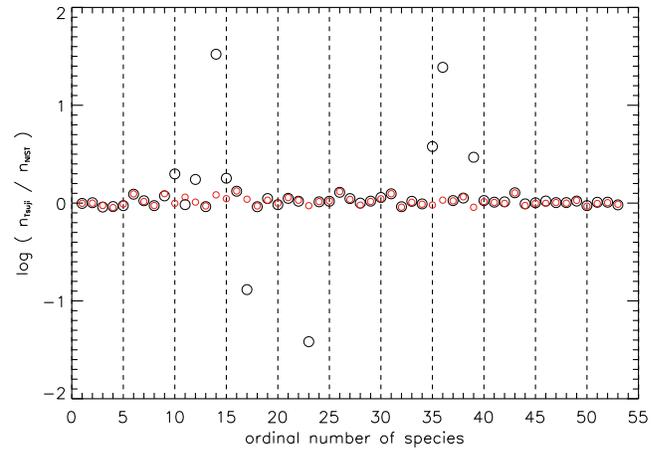}}
\caption{
The ratio between Tsuji (with $\Delta G^{\circ}$ for CS replaced by the value from NIST) and
NIST equilibrium concentrations at conditions specified in Sect.\,\ref{TEWC98}
-- black circles. Red circles mark the ratio, with additional replacement of
$\Delta G^{\circ}$ for 
CS, CN, SiC$_2$, SiH, C$_3$, CH$_3$, CH$_2$, SiN, NH$_2$, HCO, and NS
in the Tsuji database by the values from WC98.
The order of species is same as in Fig.\,\ref{err_nist}}
\label{err_tsall}
\end{figure}
\begin{figure}[!bt]
\centering
\resizebox{\hsize}{!}{\includegraphics[angle=90]{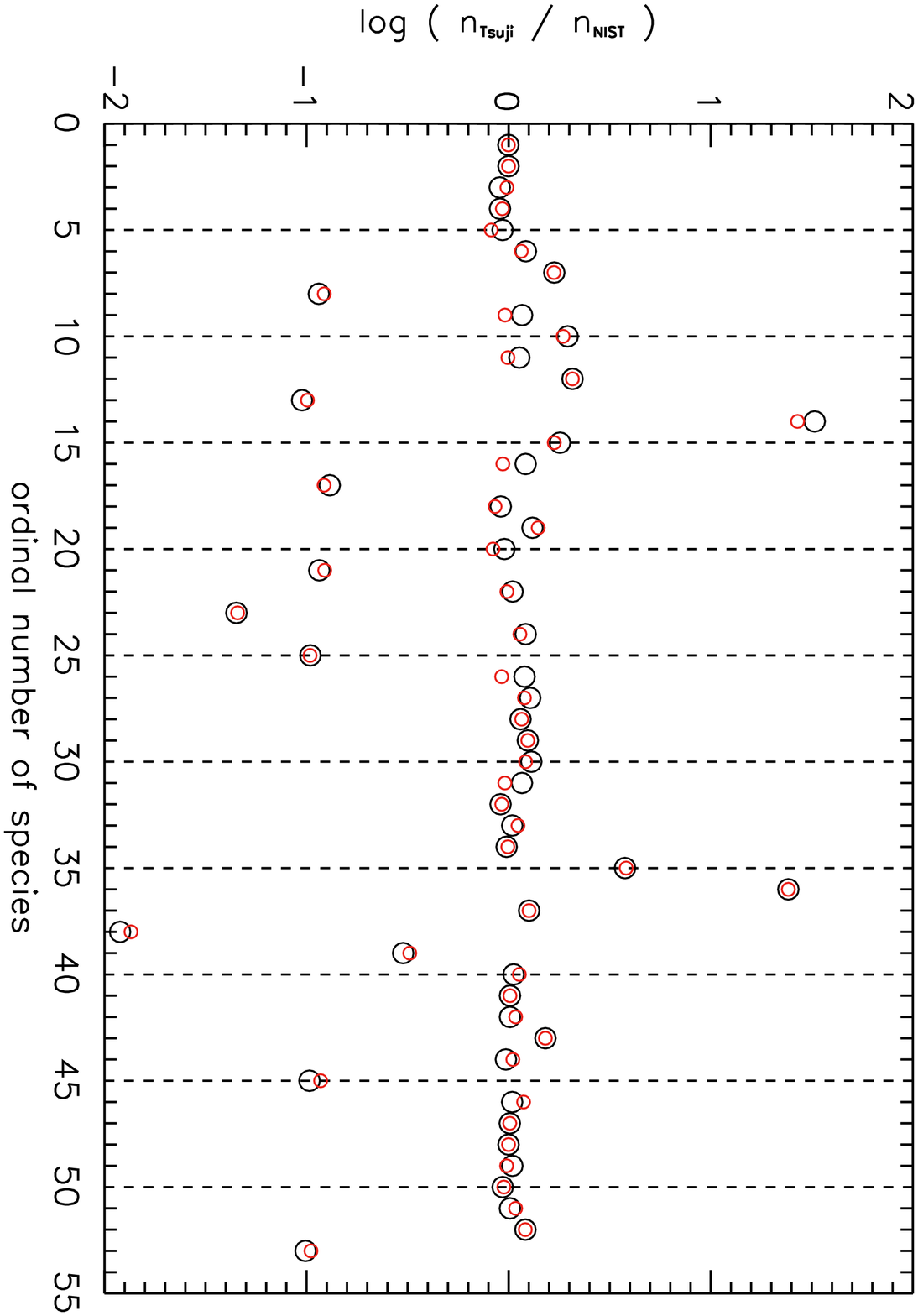}}
\caption{The ratio between Tsuji and NIST equilibrium concentrations at conditions 
specified in Sect.\,\ref{TEWC98} -- black circles. Red circles mark the ratio, when
$\Delta G^{\circ}$ for C$_2$H$_2$ in the Tsuji database was replaced by the value from NIST.
The order of species is same as in Fig.\,\ref{err_nist}}
\label{err_tsC2H2}
\end{figure}
Another~widely used database, especially in the modelling of stellar atmospheres, is the set of
dissociation constants compiled by \cite{T73}. The thermal dependence of
$K_{\rm p}$'s in this database is given by:
\begin{equation}
log K_{\rm p} = c_0 + c_1 \theta + c_2 \theta^2 + c_3 \theta^3 + c_4 \theta^4,
\label{kpT}
\end{equation}
here: $c_{\rm i, i=0,1,...,4}$ - are coefficients derived by Tsuji
and~$\theta=\frac{5040}{T}$. This equation allow us to compute dissociation
constants at the temperature of interest (T\,=\,2062\,K). The corresponding
$\Delta G^{\circ}$'s, computed from Eq.\,\ref{deltaG}, are collected in
Cols. 5 and 10 of Table~\ref{stale}, and compared with WC98 and NIST values 
in Fig.\,\ref{err_gibbs} (pluses). The Tsuji`s database does not contain 
seven molecules among 46 considered: C$_4$H$_2$ -- \#16, 
HCS -- \# 25,  C$_4$H -- \#26, HCSi -- \#27, HNSi -- \#28,
SiCH$_2$ -- \#30, and , C$_3$H$_2$ -- \#31 (empty spaces in 
Table~\ref{stale}, and in Fig.\,\ref{err_gibbs}).
In general, agreement between $\Delta G^{\circ}$'s from Tsuji and WC98 
is worse than that in case of NIST. The largest differences are seen 
for: CS -- \#7, CN -- \#10, SiC$_2$ -- \#11, SiH -- \#12,  C$_3$ -- \#13,  
CH$_3$ -- \#15, CH$_2$ -- \#17, SiN -- \#23, NH$_2$ -- \#35, HCO -- \#36, 
and NS -- \#39. $\Delta G^{\circ}$'s for other molecules differ less than 
about 4 [kJ/mol]. 

The influence of differences between the Tsuji and NIST $\Delta G^{\circ}$'s on species
concentration can be seen in Fig.\,\ref{err_tsCS}\footnote{\label{online} Figs.3--5 
and 7 are available only in the online version.}
 -- black circles.
The order of species is same as in Fig.\,\ref{err_nist}.
We plotted there the logarithm of Tsuji concentrations relative to
those of NIST (using the WC98 values of $\Delta G^{\circ}$ in case of 
molecules with missing data). The consistency is rather 
poor. Concentrations for all molecules with considerable discrepancies in 
$\Delta G^{\circ}$'s are significantly different, but additionally some other 
species reached well different equilibrium state.
For example, for all sulfur-bearing species, except of CS -- \#7 (i.e. SiS -- \#8, 
HS -- \#14, H$_2$S -- \#21, HCS -- \#25, S$_2$ -- \#38, NS -- \#39, 
SO -- \#45,  S -- \#53), we see very large discrepancies (even by factor of 
84! for S$_2$), in spite of the fact that differences between 
$\Delta G^{\circ}$'s are significant only in case of CS and NS. Among other 
molecules with significant discrepancy in concentrations are those containing 
C:  C$_3$ -- \#13,   CH$_2$ -- \#17 and  HCO -- \#36; two silicon-bearing 
molecules: SiN -- \#23 and SiH -- \#12; and 
radical NH$_2$ -- \#35.

\subsection{The influence of changes in the $\Delta G^{\circ}$'s on the TE concentrations}

Here we test the influence of changes in $\Delta G^{\circ}$ on equilibrium molecular 
content of examined gas.
\begin{figure*}[!bt]
\centering
\resizebox{\hsize}{!}{\includegraphics[angle=90]{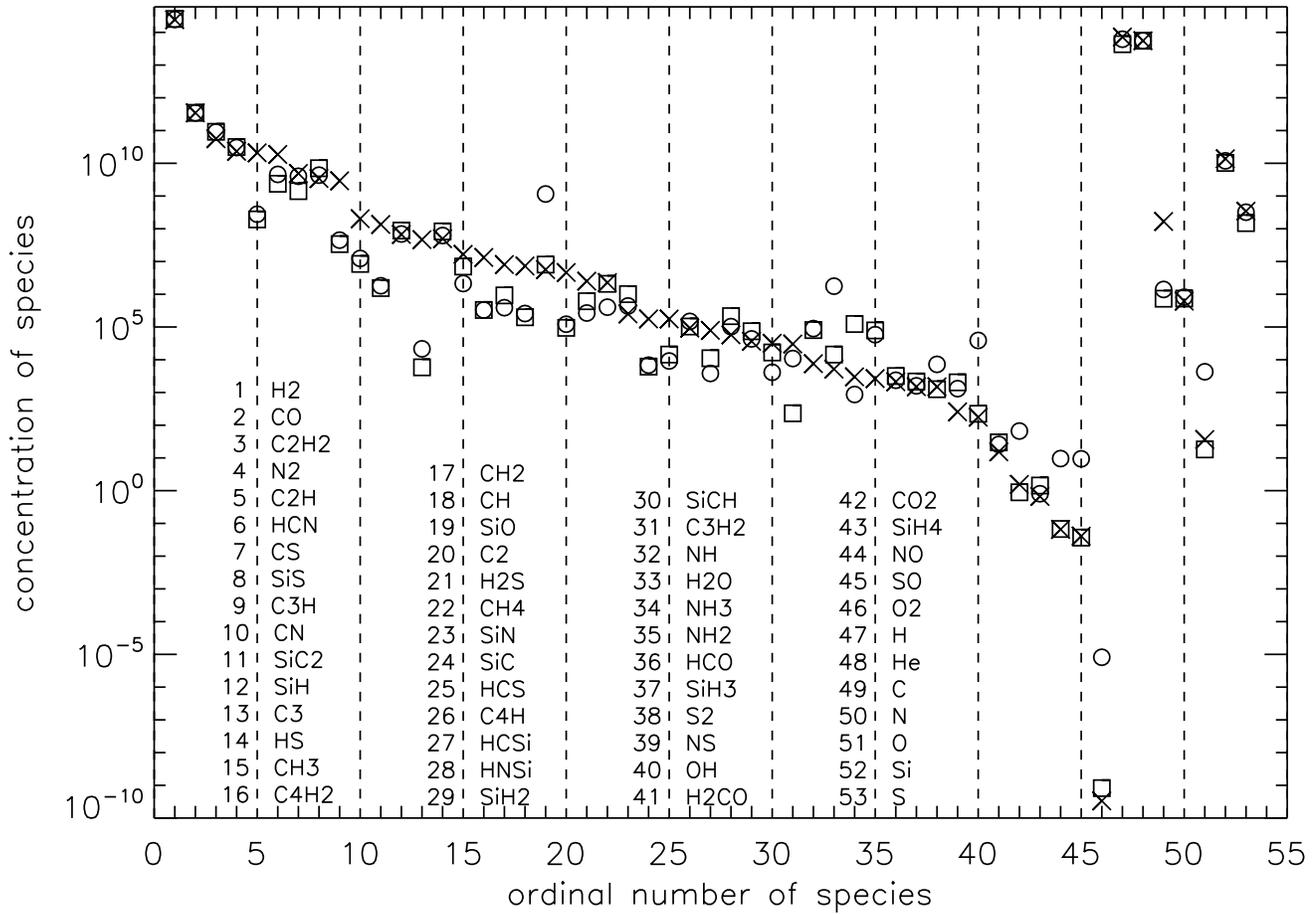}}
\caption{ Comparison between species concentrations in [cm$^{-3}$]
derived by our kinetic code for case 1 -- crosses, 
and for case 2 -- circles, at conditions specified in the 1$^{st}$ 
entry of Table\,2 in WC98. Squares mark concentration for case 2, 
with rate constants updated using the UDFA05 database. The order of 
species is same as in Fig.\,\ref{err_nist}.}
\label{net_all_compare}
\end{figure*}
\begin{figure*}[!ht]
\centering
\includegraphics[width=16.3cm, height=19cm, angle=0]{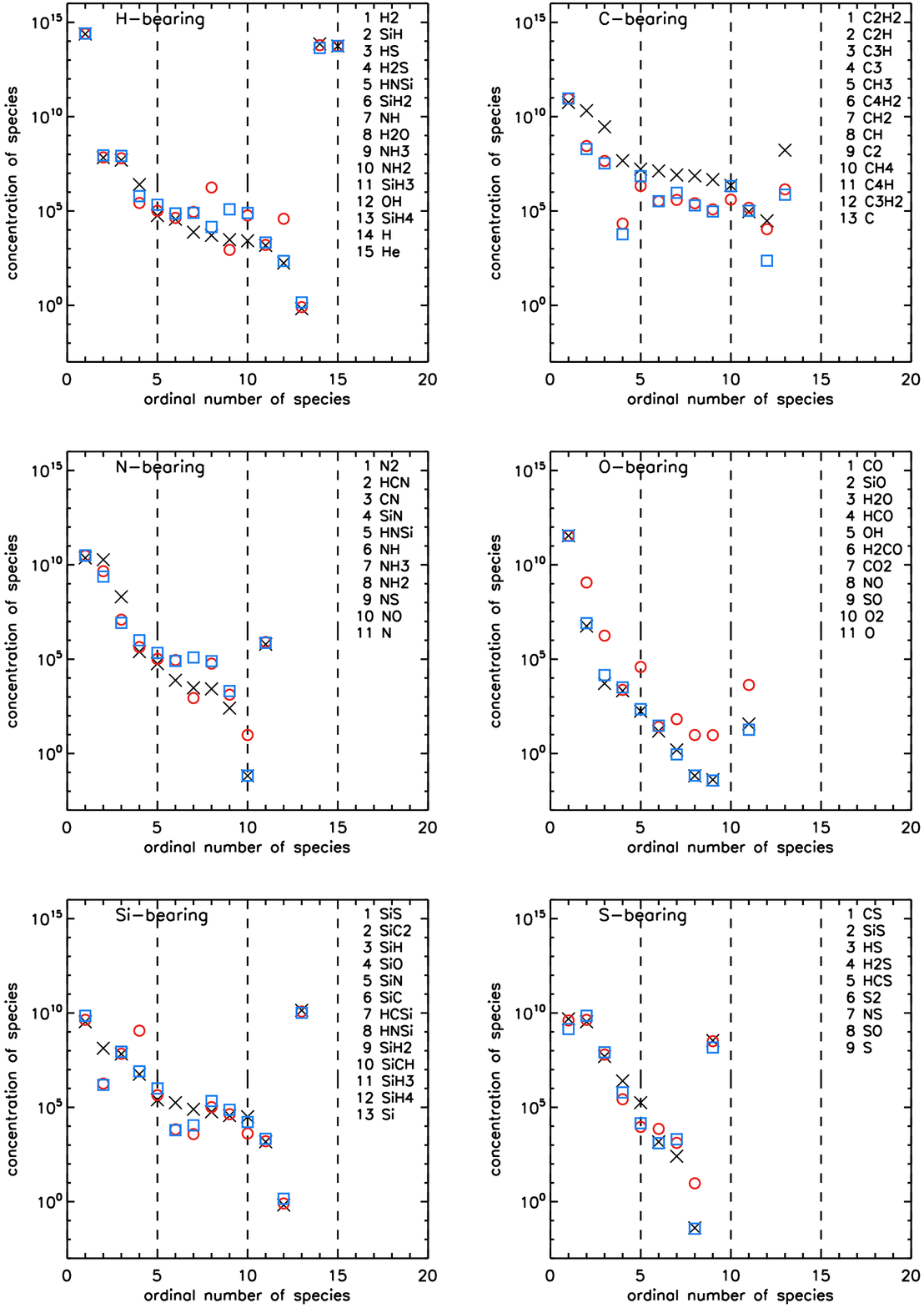}
\caption{The network balance. Comparison of species concentrations in [cm$^{-3}$] 
computed for modified WC98 network (case 1 - crosses), which are consistent
with TE solution, and for WC98 network (case 2 - circles). 
Squares mark case similar to case 2,  with rate constants updated using data 
from the UDFA05 database. The concentrations are splitted into H-bearing, 
C-bearing, N-bearing, O-bearing, Si-bearing, and S-bearing molecules. For each 
panel we defined the separate ordinal number of species according to decreasing 
equilibrium concentrations with exception of atom(s) which are shown on the right 
side of each panel.}
\label{WC98net}
\end{figure*}

In the first test we only replaced the Tsuji $\Delta G^{\circ}$ for CS (this molecule 
has the largest equilibrium concentration among S-bearing molecules) by the smaller
value from the NIST database. 
The correlation between the Tsuji and NIST concentrations becomes much better
(see red circles in Fig.\,\ref{err_tsCS}$^{\ref{online}}$), especially for family of all S-bearing 
molecules. Decrease in concentration of CS -- \#7 releases atomic carbon and 
sulfur, and the excess of sulfur increases concentrations of all S-bearing 
species. However at the same time, the increase in SiS -- \#8 concentration cause the 
decrease of concentration for all other silicon-bearing species, even for 
carbon-bearing ones: SiC$_2$ -- \#11 and SiC -- \#24. Therefore, it seems that SiS 
molecule controls the silicon chemistry in the examined case. The considerable 
increase in concentration of sulfur monoxide (SO -- \#45) is caused by both: an
access to sulfur due to decrease of $n_{\rm CS}$ and an access to oxygen due 
to decrease of $n_{\rm SiO}$ (SiO -- \#19). Furthermore, enrichment of 
the gas in carbon results in increasing concentrations of all molecules composed
of only C or carbon and hydrogen. Note, that decreasing of $\Delta G^{\circ}$ for 
NS - \#39 (molecule with the second largest difference between $\Delta G^{\circ}$
in the NIST and Tsuji databases) would result in much better agreement between 
corresponding concentrations {\it only} for this molecule).

As a starting point for the second test we chosen concentrations marked by the 
red circles in Fig.\,\ref{err_tsCS}$^{\ref{online}}$.
They are repeated now by black circles 
in Fig.\,\ref{err_tsall}$^{\ref{online}}$. The order of species is same as in Fig.\,\ref{err_nist}.
Red circles in Fig.\,\ref{err_tsall}$^{\ref{online}}$ show now the 
Tsuji concentrations relative to the NIST ones obtained after additional replacement of 
$\Delta G^{\circ}$'s by the NIST values for all molecules, which show significant
discrepancy in the Gibb's energies (i.e. for CN - \#10, SiC$_2$ - \#11, SiH - \#12, 
C$_3$ - \#13, CH$_3$ - \#15, CH$_2$ - \#17, SiN - \#23, NH$_2$ - \#35, HCO - \#36, 
and NS - \#39 -- see Fig.\,\ref{err_gibbs} and Table\,\ref{stale}). As we can see, changes in 
$\Delta G^{\circ}$'s for these molecules alter almost only concentrations for these species.
The exception is C$_3$, where atomic carbon released by its decreasing abundance (due to the
smaller $\Delta G^{\circ}$ in the NIST database) is bound also 
into other carbon-bearing molecules: C$_3$H -- \#9, 
SiC$_2$ --\#11, C$_4$H$_2$ -- \#16, C$_4$H -- \#26, and C$_3$H$_2$ -- \#31.

Finally, we present results, which show how relatively small changes of 
$\Delta G^{\circ}$ for abundant molecule like acetylene (C$_2$H$_2$ -- \#3) 
alter the equilibrium 
abundances of other molecules. $\Delta G^{\circ}$ for acetylene from the Tsuji 
database is {\it only} about 3.6 [kJ/mol] smaller than that from NIST (see 
Table\,\ref{stale}). In Fig.\,\ref{err_tsC2H2}$^{\ref{online}}$ we present the ratio between 
the Tsuji and NIST equilibrium concentrations by black circles, while red 
circles show these ratios computed with $\Delta G^{\circ}$ replaced only for acetylene 
by the larger value from the NIST. As we can see, even so small change in 
$\Delta G^{\circ}$ for this parent and abundant molecule alter significantly the obtained 
concentrations for almost all other species. With larger $\Delta G^{\circ}$ 
for acetylene more carbon 
is used for formation of this molecule and therefore concentration of other 
C-bearing molecules are significantly decreased. This affects also
concentrations of other species. In case of silicon-bearing molecules all silicon
released from SiC$_2$ is bounded into SiS and SiO. The less sensitive to the 
change of $\Delta G^{\circ}$ for acetylene turned out to be such molecules 
as: SiH -- \#12, HCS -- \#25, SiH$_2$ -- \#29, HCO -- \#36, SiH$_3$ -- \#37, 
H$_2$CO -- \#41, and SiH$_4$ -- \#43.

\section{Test of chemical codes: The chemical kinetics}

In the outer part of the stellar atmospheres one can expect that non-TE effects 
are becoming more and more important and non-TE methods to determine properly 
concentrations of chemical species are necessary.
We have developed a kinetic code, which can treat time-dependent chemistry 
and describe here the performed tests. Now, however, instead of using 
dissociation constants (or equivalently $\Delta G^{\circ}$'s)
we need to consider an appropriate reaction network.
In this work we consider reaction network of WC98.

\subsection{The kinetic code and reaction network}

To find molecular concentrations we have to consider all chemical reactions 
from the given network, which lead to formation and destruction of all molecules 
under study. The evolution of concentration with time for 
species X can be written as:
\begin{equation}
\frac{d n_{\rm X}}{d t} = F_{\rm X} - D_{\rm X},
\label{continuity}
\end{equation}
where: the F$_{\rm X}$ stands for reactions responsible for production
of species X, and $D_{\rm X}$ for its destruction.
In the developed code the rate equations, defined for each considered reaction, 
are accumulated. The rate equation describes the rate 
of disappearance of each of the reactants and on the other
hand the appearance rate of each of the products. For example, for reaction
A + B $\rightarrow$ C + D the rate equation is defined as:
\begin{equation}
-\frac{dn_{\rm A}}{dt}=-\frac{dn_{\rm B}}{dt}=\frac{dn_{\rm C}}{dt}=
\frac{dn_{\rm D}}{dt} = k \, n^{\gamma}_{\rm A}\,n^{\delta}_{\rm B},
\label{rate}
\end{equation}
where: $n_{\rm A}$, $n_{\rm B}$ -- are the concentrations of reactants,
$n_{\rm C}$, $n_{\rm D}$ -- are the concentrations of products, and
the parameters $\gamma$ and $\delta$ characterise order of reaction with 
respect to each reactant. These parameters are derived experimentally and 
their sum determines the reaction order. The coefficient $k$ in the above
equation is known as a reaction rate constant and determines reaction 
dependence on the temperature. In our model the order of considered reactions 
was always the same as the reactants number, i.e. $\gamma$ and $\delta$ 
were equal 1.

We search for the steady state solution of the above set of rate equations~
using~{\textsf DVODE}~solver for systems of ordinary differential equation 
written at Lawrence Livermore National 
Laboratory\footnote{http:://www.llnl.gov/CASC/odepack/}.
{\textsf DVODE} solves the stiff system of differential equations of first order, just 
like systems which are often met in the chemical kinetics. The solver must be 
updated with the subroutine providing the net reaction rates and eventually 
subroutine providing Jacobian of the system. These two routines are constructed 
automatically by searching the network for given molecules and preparing set of 
corresponding rate equations for given range of temperature.
The WC98 reaction network used in our study includes 53 species (7 elements and 
46 molecules -- given in Table\,\ref{elemAbun} and Table\,\ref{stale}, 
respectively). The molecules are composed of six elements: H, C, N, O, Si, 
and S. In Table 5 of WC98 there are data for 352 reactions. 
Among them there are 139 reactions in both directions (hereafter
"two-way") and 213 reactions which do 
not have the backward reaction documented (hereafter "one-way", 
marked by " $\leftrightarrow$" in their table).  
 However there are some miss printings in this table:
\begin{enumerate}
\item 
The "two-way" reaction number 53 
(H$_2$ + N$_2$ $\rightarrow$ NH + NH) has rate constant documented as being zero. 
We computed this missing rate constant (by method described below - see Eqs.\,\ref{kb} and 
\ref{kr}) using 
equilibrium constant and the rate constant for its backward reaction (i.e. for reaction number 
269: NH + NH $\rightarrow$ H$_2$ + N$_2$). Note, however, that such approach causes that 
reaction 269 must be treated as "one-way" process. After this change number of "one-way" 
reactions in network increased by one to 214, while number of "two-way" reactions decreased
to 137; 
\item The reaction number 78 (C + NS $\rightarrow$ CS + N) is indicated  as a "two--way" 
reaction (there is no " $\leftrightarrow$" in the table), but this reaction does not have 
backward reaction documented and, in fact, should be traeated as "one--way" process. 
This change caused that number of "one-way" reactions increased to 215, while the number of
"two-way" reactions decreased to 136;
\item
The reaction number 217 (S+N$_2$ $\rightarrow$ S + NS) is incorrect. It seems that the correct 
version of this reaction is: S+N$_2$ $\rightarrow$ N + NS. However, such reaction is already 
present in the network, therefore we simply deleted reaction number 217 from the list.
This way, number of "one-way" reactions decreased by one to 214;
\item
The coorect version of reaction 279 (SiH + NO $\rightarrow$ SiO + H) should be 
SiH + NO $\rightarrow$ SiO + NH. This does not change the statistics of rections.
\end{enumerate}
In summary, we have 564 reactions in the network according to Table 5 of WC98: 214 in "one-way" 
+ 214 backward -- determined as described below + 136 "two-way" reactions.

The rate constants for forward and backward reactions documented in Table\,5 
of WC98 were computed via equation in Arrhenius form:
\begin{equation}
k_{\rm f,b} = A \left( \frac{T}{300} \right) ^{\beta} \exp{\left( -\frac{E_{\rm a}}{T} \right)},
\label{kfb}
\end{equation}
where: $k_{\rm f}$ and $k_{\rm b}$ are the rate constants of forward and 
backward reactions, respectively, $A$, $\beta$, and $E_{\rm a}$ are the 
Arrhenius parameters given in that table. In case of missing backward reactions, 
WC98 calculated the backward rate constant $k_{\rm b}$ from the 
thermodynamics of the reaction (see Eq.\,4 in WC98).
In our computations the backward rate constant $k_{\rm b}$ for specific reaction and given 
temperature was derived from the ratio of rate constant for forward reaction,
and the reaction equilibrium constant $K_{\rm r}$, i.e.:
\begin{equation}
k_{\rm b}=k_{\rm f}/K_{\rm r}.
\label{kb}
\end{equation}
On the other hand, the equilibrium constant for given reaction can be obtained 
from the ratio of equilibrium concentrations for products and reactants. For 
example, for reaction AB~+~C~$\rightarrow$~AC~+~B the corresponding 
equilibrium constant is given by:
\begin{equation}
K_{\rm r}=\frac{n_{\rm AC} n_{\rm B}}{n_{\rm AB} n_{\rm C}},
\label{kr}
\end{equation}
where: $n_{\rm AB}$, $n_{\rm C}$, $n_{\rm AC}$, $n_{\rm B}$ -- equilibrium
concentrations of reactants and products, respectively.
Therefore after inserting $K_r$ given by Eq.~(\ref{kr}) into Eq.\,({\ref{kb}})
we have:
\begin{equation}
k_{\rm b}=k_{\rm f} \frac{n_{\rm AB} n_{\rm C}}{n_{\rm AC} n_{\rm B}}.
\label{kback}
\end{equation}
Thus, by computing {\it missing} backward rate constants, we have modified the 
original network of WC98 and use it to test our kinetic code.

\subsection{Comparison of TE concentrations derived with WC98 and UDFA05 reaction networks}

The chemical timescales in the dense and warm part of the envelope
are very short, so the concentrations of species should reach the
steady state values very quickly. Therefore, to test our kinetic code we decided
to reproduce the equilibrium concentrations of the 53 species discussed above.
In such case, as an input to the code we need the initial concentrations of
elements only (values are taken from Table\,\ref{elemAbun}), the total gas 
density and its temperature (again we have adopted physical conditions specified 
in the 1$^{st}$ entry of Table\,2 in WC98).

To test our chemical kinetic code we have adopted rate constants 
only for the forward reactions\footnote{As a
forward reaction, among reactions with rate constants given for
both directions, we chose the reaction with more precise rate constant
i.e. first when Arrhenius parameter $\beta$ (see Eq.~\ref{kfb}) is
different from zero. If $\beta$\,=\,0 for both reactions, we took reaction
with smaller Arrhenius parameter $E_{\rm a}$. In case of reactions without the 
backward rate constants we treat all of them as a forward reactions.} 
(282\,=\,136/2+214 -- the network of WC98 requires the rate equations for 
564 reactions)  and we have computed the rate constants for all (282) backward reactions
according to the method described above (hereafter case 1).  
 As expected, in this case our code reproduces the WC98 equilibrium concentrations.
The results are marked as crosses in Fig.\,\ref{net_all_compare}. The species 
(except atoms collected at the right side of the plot) are plotted in order 
of decreasing abundances. In addition, we have tested stability of our code 
by assuming that initial
chemical composition of all species is equal to that at the thermodynamical equilibrium 
for the considered conditions. No changes in the species concentrations were 
seen after very long time (10$^8$ sec). Therefore, for further computations 
we used the equilibrium concentrations as an input to our
code (for its faster convergence). Note that the results are the same when we 
start with a purely atomic gas.

In the next test (hereafter case 2) we took all rate constants available in
the WC98 network (i.e. 350\,=214\,+\,136 - see the previous subsection) and computed 
the rate constants only for the missing (214) backward reactions by using NIST 
thermochemical data (in fact, this approach is the same as described by WC98).
The results are presented in Fig.\,\ref{net_all_compare} as circles.
It is seen, that the present steady state differs from the thermodynamical equilibrium 
concentrations (case 1). 
More detailed comparison for all species, which belong to a given family (molecules 
containing element: H, C, N, O, Si, and S, respectively) are presented in six panels 
in Fig.\,\ref{WC98net}$^{\ref{online}}$ as circles: going from the left to right and from the 
top to bottom for: H, C, N, O, Si, and S. For each panel we defined the separate ordinal number
of species according to decreasing equilibrium concentrations with exception of atom(s) which
are shown on the right side of each panel.

The inconsistency in species abundances is rather high and only fifteen species
differ less then 30 \% (i.e.
He, CO, H$_2$, SiH, SiH$_3$, S, HCO, Si, H, CS, SiH$_2$, SiH$_4$, HS, SiS, N$_2$, and N).     
The highest differences are seen for oxygen-based molecules. Only
formyl (HCO) and the most abundant among O-bearing molecules -- carbon monoxide (CO) 
are in relatively good agreement 
with the corresponding equilibrium concentrations. Concentrations of other molecules 
(i.e. SiO, H$_2$O, OH, H$_2$CO, CO$_2$, NO, SO, and O$_2$) are far from equilibrium values.

The WC98 network was created on the basis of the RATE95 database 
\citep{M97}.  The discrepancies present in the above results
encourage us to update the rate constants using the UDFA05 database 
\citep{W06}. We found that 162 reactions among 564  
have now the new values of the rate constants.
We performed the same test as in case 2
(i.e. only missing backward rates were computed using Eq.\,\ref{kback})
with the reaction network updated using the UDFA05 database. 
The results are presented as squares in Fig.\,\ref{net_all_compare}, and
the family splitting in this case is shown in
Fig.\,\ref{WC98net}$^{\ref{online}}$ also by squares.

The agreement with the equilibrium state for oxygen-bearing molecules now is 
much better.  However, the discrepancies in abundances of carbon-bearing 
molecules are still present. Moreover, the differences with equilibrium results 
for nitrogen-bearing molecules are even higher.

The most likely source of inconsistency is the problem with the determination of 
reaction rate constants for the silicon- and sulfur-bearing species.
WC98 estimated the rate constants for these reactions from similar reactions 
involving covalent species, i.e. oxygen for sulfur and carbon for silicon (see comments 
in Table 5 of WC98). When this group of reactions was excluded from the
studied network, the abundances of carbon-bearing and oxygen-bearing molecules 
became more consistent with LTE results. However, the results for 
nitrogen-bearing species are still unsatisfactory and additional work on the chemical 
network extension is needed.

\section{Conclusions}

In this paper we have performed tests of our chemical (equilibrium and kinetic) 
codes with the aim to use them during self-consistent modelling of dynamics and 
chemistry in outflows from C-rich AGB stars.
 
The LTE molecular distribution at the base of the wind is needed as the boundary 
condition to determine chemical composition in the circumstellar envelope of the 
AGB star. In such LTE calculations it is necessary to keep in mind the possible 
influence of the different sets of equilibrium dissociation constants on
the species concentration. Analysis of the conducted test calculations shows that, 
fortunately, such influence for not abundant molecules is rather small. However, 
for more abundant species the influence is crucial and the used sets of equilibrium 
dissociation constants should  be carefully selected and checked. 
In particular, our calculations have shown that the NIST database 
reproduce well the WC98 equilibrium concentrations, while agreement 
between WC98 and Tsuji is much worse.

Steady state solution obtained with kinetic code for WC98 reaction network
is different from the thermodynamical equilibrium solution.  This would affect the full 
time-dependent radiative-hydrodynamic computations of the inner wind. Note that 
CN and C$_2$ molecules, important for the opacity computations in the atmosphere of carbon 
rich star, are underabundant by more than one order of magnitude 
in comparison to equilibrium concentration (see Fig.\ref{net_all_compare}). 
Our kinetic computations also show the strong overabundance of oxygen-bearing molecules
(especially of SiO, H$_{2}$O, and OH) in comparison to the LTE approach. It would be 
interesting to investigate how these overabundances influenced the final results 
of the inner wind studies by \citet{W98} and \citet{C06}.

To make both, LTE and kinetic, models consistent we propose to replace all 
reaction rates in backward direction\footnote{see footnote 6} by reaction rates 
computed from forward reactions with usage of thermochemical data. Then,
in the limit of high density and temperature (i.e. when chemical timescales are 
much shorter than dynamical timescales) the kinetic steady state solution 
approach the local thermodynamical equilibrium.

Observations \citep[]{T97} and the astrochemical modelling
\citep[WC98,][]{C06} 
are showing that the base of inner wind in AGBs is a region of active chemistry.
Therefore, chemical network, which may be used to investigate the composition
of this region should include extended set of chemical reactions such as 
neutral--neutral involving metals and photochemistry induced by the stellar 
radiation field. This will be direction of our future studies.

\begin{acknowledgements}
We are very indebted to the referee Karen Willacy for providing us with
the original Gibb's free energies used in WC98 and for comments which allowed
us to improve the manuscript.

This work has been partly supported by grants 2.P03D 017.25 and 
1.P03D.010.29 of the Polish State Committee for Scientific Research.
\end{acknowledgements}

\bibliographystyle{aa}
\bibliography{pulecka1.bib}

\end{document}